\documentstyle[sprocl]{article}

\def\be{\begin{equation}}
\def\ee{\end{equation}}
\def\bea{\begin{eqnarray}}
\def\eea{\end{eqnarray}}

\newcommand{\nn}{\nonumber}

\newcommand{\N}{\mbox{\rm I$\!$N} }
\newcommand{\R}{\mbox{\rm I$\!$R} }
\def\C{\mbox{\rm {I\kern-.520em C}}}

\newcommand{\sign}{\mbox{\rm sign} }
\newcommand{\e}{ \mbox{\rm e} }

\newcommand{\p}{\partial}
\newcommand{\btd}{\bigtriangledown}
\newcommand{\btu}{\bigtriangleup}
\newcommand{\tri}{\Delta}

\newcommand{\sums}{\sum\limits}
\newcommand{\const}{\mathop{\rm const}\nolimits}

\begin{document}
\title{TODA CHAINS WITH TYPE $A_m$ LIE ALGEBRA FOR MULTIDIMENSIONAL
CLASSICAL COSMOLOGY WITH INTERSECTING $p$-BRANES}
\author{V. R. GAVRILOV, V. N. MELNIKOV}
\address{Centre for Gravitation and Fundamental Metrology
VNIIMS,\\3-1 M. Ulyanovoy St., Moscow 117313, Russia\\
E-mail:melnikov@rgs.phys.msu.su}

\maketitle\abstracts{ We consider a $D$-dimensional cosmological model  
describing an evolution of $(n+1)$ Einstein factor spaces ($n\geq 2$) in 
the theory with several dilatonic scalar fields and generalized 
electro-magnetic forms, admitting an interpretation in terms of 
intersecting $p$-branes.  The equations of motion of the model are 
reduced to the Euler-Lagrange equations for the so called 
pseudo-Euclidean Toda-like system. We consider the case, when 
characteristic vectors of the model, related to $p$-branes configuration 
and their couplings to the dilatonic fields, may be interpreted as the 
root vectors of a Lie algebra of the type $A_m$. The model is reduced to 
the open Toda chain and integrated. The exact solution is presented in the 
Kasner-like form.}

\section{Introduction}
Last years have witnessed a growth of interest to classical $p$-brane 
solutions of (bosonic sector of) supergravities in various dimensions.
\cite{St}$^-$\cite{GrIM}
This interest is inspired by a conjecture that $D=11$ supergravity is a 
low-energy effective field theory of eleven-dimensional fundamental 
$M$-theory, 
which (together with so called $F$-theory) is a candidate for unification 
of five known $D=10$ superstring theories. 
\cite{DKL}$^-$\cite{HV}  Classical $p$-brane solutions 
may be considered as a method for investigation of interlinks between 
superstrings and $M$-theory. 

In this paper we consider a generalized bosonic sector of early 
supergravity theories \cite{CJS} in the form of multidimensional 
gravitational model with several dilatonic fields and Maxwell-like forms 
of various ranks.  When $D$-dimensional space-time manifold is a product 
of several Einstein factor spaces, the most convenient to use 
$\sigma$-model approach (see, for instance, 
\cite{IM4}$^,\,$\cite{RZ96}$^,\,$\cite{R98}) 
for constructing  exact solutions with $p$-branes.
It was shown \cite{IMcosm}
that for cosmological space-times the equations of motion to such 
$\sigma$-model
are reduced to the Euler-Lagrange equations for a pseudo-Euclidean 
Toda-like Lagrange system. The methods for integrating of pseudo-Euclidean 
Toda-like systems were developed in the papers (see, for instance, 
\cite{IM3}$^-$\cite{GIM} and references therein) devoted to the 
multidimensional cosmologies with milticomponent perfect fluid.  Here we 
integrate the intersecting $p$-branes model reducible to Toda chain 
associated with Lie algebra of the type $A_m$.
                    
\section{\bf The general model}
\setcounter{equation}{0}

Following the papers 
\cite{IM4}$^,\,$\cite{IM41}$^,\,$\cite{IM}$^-$\cite{Br}$^,\,$\cite{IMcosm}$^-$\cite{GrIM}
we study a classical
model in $D$ dimensions described by the action 
\bea\label{1}
S =&& \int_{\cal M} d^{D}z \sqrt{|g|} 
\{ {R}[g] - h_{\alpha\beta}\; g^{MN} \partial_{M} 
\varphi^\alpha \partial_{N} \varphi^\beta \\ \nn && - \sum_{a \in \Delta} 
\frac{1}{n_a!} \exp[ 2 \lambda_{a} (\varphi) ] (F^a)^2_g \},
\eea
where $g = g_{MN} dz^{M} \otimes dz^{N}$ is the metric
on $D$-dimensional manifold  ${\cal M}$
($M,N =1, \ldots, D$), $|g| = |\det (g_{MN})|$.
We denoted by $\varphi=(\varphi^\alpha)\in \R^l$
a vector from dilatonic scalar fields,
$(h_{\alpha\beta})$ is a non-degenerate $l\times l$ matrix ($l\in \N$)
and $\lambda_{a}$ is a $1$-form
on $\R^l$: $\lambda_{a} (\varphi) =\lambda_{a \alpha} \varphi^\alpha$,
$a \in \Delta$, $\alpha=1,\ldots,l$. 
\be
F^a =  dA^a =
\frac{1}{n_a!} F^a_{M_1 \ldots M_{n_a}}
dz^{M_1} \wedge \ldots \wedge dz^{M_{n_a}}
\ee
is a $n_a$-form ($n_a \geq 1$) on ${\cal M}$.
In Eq.~(\ref{1}) we denoted
\be
(F^a)^2_g =
F^a_{M_1 \ldots M_{n_a}} F^a_{N_1 \ldots N_{n_a}}
g^{M_1 N_1} \ldots g^{M_{n_a} N_{n_a}},
\ee
$a \in \Delta$, where $\Delta$ is some finite set.

Equations of motion corresponding to (\ref{1}) have the following
form
\bea\label{4}
R_{MN} - \frac{1}{2} g_{MN} R  =   T_{MN},
\\
\label{5}
{\btu}[g] \varphi^\alpha -
\sum_{a \in \Delta}\frac{\lambda^{\alpha}_a}{n_a!}
e^{2 \lambda_{a}(\varphi)} (F^a)^2_g = 0,
\\
\label{6}
\nabla_{M_1}[g] (e^{2 \lambda_{a}(\varphi)}
F^{a, M_1 \ldots M_{n_a}})  =  0,
\eea
$a \in \Delta$; $\alpha=1,\ldots,l$.
In Eq.~(\ref{5}) $\lambda^{\alpha}_{a} = h^{\alpha \beta}
\lambda_{a \beta }$, where $(h^{\alpha \beta})$
is a matrix inverse to $(h_{\alpha \beta})$.
In (\ref{4}) we denoted
\bea
T_{MN} =   T_{MN}[\varphi,g]
+ \sum_{a\in\Delta} e^{2 \lambda_{a}(\varphi)} T_{MN}[F^a,g],
\eea
where
\bea
T_{MN}[\varphi,g] =
h_{\alpha\beta}\left(\p_{M} \varphi^\alpha \p_{N} \varphi^\beta -
\frac{1}{2} g_{MN} \p_{P} \varphi^\alpha \p^{P} \varphi^\beta\right),
\\
T_{MN}[F^a,g] = \frac{1}{n_{a}!}\left[ - \frac{1}{2} g_{MN} (F^{a})^{2}_{g}
+ n_{a}  F^{a}_{M M_2 \ldots M_{n_a}} F_{N}^{a, M_2 \ldots M_{n_a}}\right].
\eea
In Eqs.~(\ref{5}), (\ref{6}) ${\btu}[g]$ and ${\btd}[g]$
are Laplace-Beltrami and covariant derivative operators
corresponding to  $g$.

We consider the manifold
\be\label{10}
{\cal M} = \R  \times M_{0} \times \ldots \times M_{n}
\ee
with the metric
\be\label{11}
g= - \e^{2{\gamma}(t)} dt \otimes dt +
\sum_{i=0}^{n} \e^{2x^i(t)} g^i ,
\ee
where $g^i  = g^i_{m_{i} n_{i}}(y_i) dy_i^{m_{i}} \otimes dy_i^{n_{i}}$
is a metric on $M_{i}$  satisfying the equation
\be
R_{m_{i}n_{i}}[g^i ] = \xi_{i} g^i_{m_{i}n_{i}},
\ee
$m_{i},n_{i}=1,\ldots,d_{i}$; $d_{i} = \dim M_i$, $\xi_i= \const$,
$i=0,\dots,n$; $n\in {\bf N}$. So all $(M_i,g^i)$ are Einstein spaces.

Each manifold $M_i$ is supposed to be oriented and connected. 
Then the volume $d_i$-form
\be\label{13}
\tau_i  = \sqrt{|g^i(y_i)|}
\ dy_i^{1} \wedge \ldots \wedge dy_i^{d_i},
\ee
and the signature
\be\label{14}
\sign \det (g^i_{m_{i}n_{i}}) = 1
\ee
are correctly defined for all $i=0,\ldots,n$.

Let
\be\label{15}
\Omega_0 = \{ \emptyset, \{ 0 \}, \{ 1 \}, \ldots, \{ n \},
\{ 0, 1 \}, \ldots, \{ 0, 1,  \ldots, n \} \}
\ee
be a set of all ordered subsets of
$I_0\equiv\{ 0, \ldots, n \}.$
For any $I = \{ i_1, \ldots, i_k \} \in \Omega_0$, $i_1 < \ldots < i_k$,
we define a volume form
\be\label{16}
\tau(I) \equiv \tau_{i_1}  \wedge \ldots \wedge \tau_{i_k}
\ee
of rank
\be\label{17}
d(I) \equiv  \sum_{i \in I} d_i = d_{i_1} + \ldots + d_{i_k}
\ee
and a corresponding $p$-brane submanifold
\be\label{18}
M_{I} \equiv M_{i_1}  \times  \ldots \times M_{i_k},
\ee
where $p=d(I)-1$ (${\rm dim M_{I}} = d(I)$).

We adopt the following "composite 
electro-magnetic" Ansatz for fields of forms
\be\label{19} 
F^a=\sum_{I\in\Omega_{a,e}}F^{(a,e,I)}+\sum_{J\in\Omega_{a,m}}F^{(a,m,J)},
\ee
where
\bea\label{20}
F^{(a,e,I)}=d\Phi^{(a,e,I)}\wedge\tau(I), \\ \label{21}
F^{(a,m,J)}=\e^{-2\lambda_a(\varphi)}
*\left(d\Phi^{(a,m,J)}\wedge\tau(J)\right),
\eea
$a\in\tri$, $I\in\Omega_{a,e}$, $J\in\Omega_{a,m}$ and
\be\label{22}
\Omega_{a,e},\Omega_{a,m}\subset \Omega_0.
\ee
(For empty $\Omega_{a,v}=\emptyset$, $v=e,m$, we put $\sums_\emptyset=0$ in
Eq.~(\ref{19})). In Eq.~(\ref{21}) $*=*[g]$ is the Hodge operator on 
$(M,g)$.

For the potentials in Eqs.~(\ref{20}), (\ref{21}) we put
\be\label{23}
\Phi^s=\Phi^s(t),
\ee
$s\in S$, where
\be\label{24}
S=S_e\sqcup S_m,  \qquad
S_v\equiv \coprod_{a\in\tri}\{a\}\times\{v\}\times\Omega_{a,v},
\ee
$v=e,m$.

For dilatonic scalar fields we also put
\be\label{25}
\varphi^\alpha=\varphi^\alpha(t),
\ee
$\alpha=1,\dots,l$.

>From  Eqs.~(\ref{20})  and (\ref{21}) we obtain 
the relations between dimensions of $p$-brane 
worldsheets and ranks of forms
\bea
d(I) = n_a - 1,  \quad I \in \Omega_{a,e},
\\ 
d(J) = D - n_a - 1,  \quad J \in \Omega_{a,m}
\eea
in electric and magnetic cases respectively.

We put the following restrictions on $\Omega_{a,v}$. Let
\be
w_1\equiv\{i \mid i\in\{1,\dots,n\},\ d_i=1\}.
\ee
The set $w_1$ describes all $1$-dimensional manifolds among $M_i$ $(i\ge0)$.
We impose the following restrictions on the sets $\Omega_{a,v}$
(\ref{22}):
\be\label{29}
W_{ij}(\Omega_{a,v})=\emptyset,
\ee
$a\in\tri$; $v=e,m$; $i,j\in w_1$, $i<j$ and
\be\label{30}
W_j^{(1)}(\Omega_{a,m},\Omega_{a,e})=\emptyset,
\ee
$a\in\tri$; $j\in w_1$. Here
\bea
W_{ij}(\Omega_*)\equiv
\{(I,J)|I,J\in\Omega_*,\ I=\{i\}\sqcup(I\cap J),\
J=\{j\}\sqcup(I\cap J)\},
\eea
$i,j\in w_1$, $i\ne j$, $\Omega_* \subset \Omega_0$ and
\be\label{32}
W_j^{(1)}(\Omega_{a,m},\Omega_{a,e}) \equiv
\{(I,J)\in\Omega_{a,m}\times\Omega_{a,e}|\bar I=\{j\}\sqcup J\},
\ee
$j\in w_1$. In (\ref{32})
\be
\bar I\equiv I_0 \setminus I
\ee
is a "dual" set. (The restrictions (\ref{29}) and (\ref{30}) are
trivially satisfied when $n_1\le1$ and $n_1=0$, respectively, where
$n_1=|w_1|$ is the number of $1$-dimensional manifolds among $M_i$).

It was shown \cite{IMcosm} that after the following gauge fixing
\begin{eqnarray}
 \gamma=\gamma_0
\equiv\sum_{i=0}^nd_ix^i,  
\end{eqnarray}
the Maxwell
equations (\ref{6}) for $s\in S_e$ and Bianchi identities ${\rm d}F^s=0$
for $ s\in S_m $ look as follows
\begin{eqnarray}
 \frac{d}{dt}
\left(
\exp[-2\gamma_0+2\chi_s\lambda_{a_s}(\varphi)]
\dot\Phi^s
\right)=0,
\end{eqnarray}
where 
\bea
\chi_s=+1, \quad v_s=e; \\ \label{2.x2}
\chi_s=-1, \quad v_s=m.
\eea
Then
\begin{eqnarray}
\dot\Phi^s=Q_s 
\exp[2\gamma_0-2\chi_s\lambda_{a_s}(\varphi)],
\end{eqnarray}
where $Q_s$ are constants, $s=(a_s,v_s,I_s)\in S$.
Let
\bea
&&Q_s\ne0, \quad s\in S_*; \\ \nn
&&Q_s=0, \quad s\in S\setminus S_*,
\eea
where $S_*\subset S$ is a non-empty subset of $S$.

For fixed $Q=(Q_s,s\in S_*)$ the equations of motion 
(\ref{4}),(\ref{5}) are equivalent (after the gauge fixing)
to the Euler-Lagrange equations 
following from the Lagrangian
\be\label{40}
L_Q=\frac12\bar G_{AB}\dot x^A\dot x^B-V_Q
\ee
under the zero-energy constraint
\begin{eqnarray}
E_Q=\frac12\bar G_{AB}\dot x^A\dot x^B+V_Q=0.
\end{eqnarray}
We denoted
\begin{eqnarray}
&&(x^A)=(x^i,\varphi^\alpha),\\
&&V_Q=-\frac{1}{2}\sum_{i=0}^n\xi^id^i\exp[-2x^i+2\gamma_0]
+\frac12\sum_{s\in S_*}Q_s^2 
\exp[2\gamma_0-2\chi_s\lambda_{a_s}(\varphi)] 
\end{eqnarray}
and
\bea
(\bar G_{AB})=\left(\begin{array}{cc}
G_{ij}&0\\
0&h_{\alpha\beta}
\end{array}\right),
\eea
\be
G_{ij}=d_i\delta_{ij}-d_id_j.
\ee
A system with Lagrangian of the type (\ref{40}) is called the 
pseudo-Euclidean Toda-like system. \cite{GIM} In the next section we 
consider a special case, when pseudo-Euclidean Toda-like system is 
reducible to the Toda lattice associated with a Lie algebra of the type 
$A_m$.

\section{Integration of the models reducible to a classical open Toda 
chain}
Now we consider the general model under the following restrictions
\begin{enumerate}
\item
The factor space $M_0$ has non-zero Ricci tensor and all remaining factor 
spaces $M_1,\ldots,M_n$ are Ricci flat, i.e.
\begin{eqnarray}
 \xi _0 \neq 0,\ \ \xi _1=\ldots=\xi _n=0.  \nn
\end{eqnarray}
\item
All nonvanishing (with $Q_s\neq 0$) $p$-branes do not "live" in the factor 
space $M_0$, i.e.
\begin{eqnarray}
0\notin I_s,\ \ \forall s\in S_*.\nn
\end{eqnarray}
\end{enumerate}

We introduce $(n+l+1)$-dimensional real vector space 
${\bf R}^{n+l+1}$ with a canonical basis $\{e_A\},\ A=0,\ldots,n+l$, where
$e_0=(1,0,\ldots,0)$ etc. Hereafter we use the following vectors
\begin{enumerate}
\item
The vector $x(t)$ we need to calculate
\begin{eqnarray}
x(t)=\sum_{A=0}^{n+l+1}x^A(t)e_A,\ \ x^A(t)=
\left(
x^i(t),\varphi^{\alpha}(t)
\right),                                   
\end{eqnarray}
$i=0,\ldots,n,\ \alpha=1,\ldots,l.$
We remind that ${\rm exp}[x^i(t)]$ is the scale factor of the 
space $M_i$, $\varphi^{\alpha}(t)$ is the $\alpha$-th   dilatonic scalar 
field.  
\item The vector $U_0$ induced by the curvature of the factor 
space $M_0$ 
\begin{eqnarray} U_0=\sum_{A=0}^{n+l+1}U^A_0 e_A,\ \ U^A_0= 
\left(
-\frac{\delta_{i0}}{d_0},0,\ldots,0
\right),               
\end{eqnarray}
$i=0,\ldots,n.$
\item
The vector  $U_s$ induced by the $p$-brane corresponding to 
\linebreak
$s=(a_s,v_s,I_s)\in S_*$
\begin{eqnarray}
U_s=\sum_{A=0}^{n+l+1}U^A_s e_A,\ \ U^A_s=
\left(
\delta_{iI_s}-\frac{d(I_s)}{D-2},-\chi_s\lambda_{a_s}^{\alpha}
\right),                                   
\end{eqnarray}
$i=0,\ldots,n,\ \alpha=1,\ldots,l.$
\end{enumerate}

Let $(.,.)$ be a symmetrical bilinear form defined in ${\bf R}^{n+l+1}$
by the following manner
\begin{eqnarray}
(e_A,e_B)=\bar G_{AB}.
\end{eqnarray}
The bilinear form $(.,.)$ for the vectors $U_0,U_s$ reads
\begin{eqnarray}
&&(U_0,U_0)=\frac{1}{d_0}-1,\ \ \  (U_0,U_s)=0,\ \forall s\in S_*,\\
&&(U_s,U_{s'})=d(I_s\cap T_{s'})-
\frac{d(T_s)d(I_{s'})}{D-2}+
\chi_s\chi_{s'}\sum_{\alpha,\beta=1}^l
\lambda_{a_s\alpha}\lambda_{a_{s'}\beta}h^{\alpha\beta}.
\end{eqnarray}
Using the notation $(.,.)$ and the vectors (46)-(48), we may present the 
\linebreak
Lagrangian and the zero energy constraint for this special type of the 
model in the form
\begin{eqnarray}\label{52}
L_Q=\frac{1}{2}(\dot x,\dot x)-V_Q,\ \ \ 
E_Q=\frac{1}{2}(\dot x,\dot x)+V_Q=0,
\end{eqnarray}
where
\begin{eqnarray}\label{53}
V_Q=a^{(0)}e^{2(U_0,x)}+\sum_{s\in S_*}a^{(s)}e^{2(U_s,x)}.
\end{eqnarray}
We used the following constants in Eq.~(\ref{53})
\begin{eqnarray}
a^{(0)}=-\frac{1}{2}\xi_0d_0,\ \ \ 
a^{(s)}=\frac{1}{2}Q^2_s.
\end{eqnarray}
Let the vectors $U_s\in {\bf R}^{n+l+1}$ satisfy the following conditions
\begin{eqnarray}
(U_s,U_s)=U^2>0,\ \forall s\in S_*,
\end{eqnarray}
\begin{equation}\label{56}
\left(C_{ss'}\right)=
\left(\frac{2(U_s,U_{s'})}{(U_{s'},U_{s'})}\right)=\left(
\begin{array}{*{6}{c}}
2&-1&0&\ldots&0&0\\
-1&2&-1&\ldots&0&0\\
0&-1&2&\ldots&0&0\\
\multicolumn{6}{c}{\dotfill}\\
0&0&0&\ldots&2&-1\\
0&0&0&\ldots&-1&2
\end{array}
\right)\quad .
\end{equation}
where $(C_{ss'})$ is the Cartan matrix for the Lie algebra 
$A_{|S_*|}=sl(|S_*|+1,C)$. We mention that the number of the vectors 
$U_0,U_s$ should not exceed the dimension of
${\bf R}^{n+l+1}$, i.e.
\begin{eqnarray}
n+l-|S_*|\equiv n'\geq 0.
\end{eqnarray}

We choose in ${\bf R}^{n+l+1}$ a basis $\{f_0,f_s,f_p\}, \ s\in S_*,
\ p=|S_*|+1,\ldots,|S_*|+n',$ with the properties
\begin{eqnarray}
&&f_0=\frac{U_0}{(U_0,U_0)},\ \ \ f_s=\frac{U_s}{(U_s,U_s)},
\ s \in S_*,\\
&&(f_0,f_p)=(f_s,f_p)=0,\ \ (f_p,f_{p'})=0, \ p\neq p',
\end{eqnarray}
$p,p'=|S_*|+1,\ldots,|S_*|+n'.$
It should be noted that if $n'=0$, then the vectors $f_p$ do not appear.

Using the decomposition
\begin{eqnarray}
x(t)=q^0(t)f_0+\sum_{s\in S_*}q^s(t)f_s+
\sum_{p=|S_*|+1}^{|S_*|+n'} q^p(t)f_p
\end{eqnarray}
with respect to this basis, we present the Lagrangian and the zero-energy 
constraint (\ref{52}) in the form
\begin{eqnarray}
L_Q=L_0+L_{Q,T}+L_f,\ \ \ E_Q=E_0+E_{Q,T}+E_f=0,
\end{eqnarray}
where
\begin{eqnarray}
&&L_0=\frac{(\dot{q}^0)^2}{2(U_0,U_0)}-V_0,
\ E_0=\frac{(\dot{q}^0)^2}{2(U_0,U_0)}+V_0,
\ V_0=a^{(0)}e^{2q^0},\\
&&L_f=E_f=\frac{1}{2}\sum_{p=|S_*|+1}^{|S_*|+n'}
(f_p,f_p)(\dot{q}^p)^2,\\
&&L_{Q,T}=\sum_{s,s'\in S_*}\frac{C_{ss'}}{4U^2}\dot{q}^s\dot{q}^{s'}-V_{Q,T}, 
E_{Q,T}=\sum_{s,s'\in S_*}\frac{C_{ss'}}{4U^2}\dot{q}^s\dot{q}^{s'}+V_{Q,T},\\
&&V_{Q,T}=\sum_{s'\in S_*}a^{(s)}{\rm exp}
\left[
\sum_{s'\in S_*}C_{ss'}q^{s'}
\right].
\end{eqnarray}
Integration of equations of motion for $q^0$ is straightforward and 
leads to the following result
\begin{eqnarray}
e^{-q^0(t)}=F(t-t_0),
\end{eqnarray}
where $t_0$ is an arbitrary constant and the function $F_0$ has the form
\begin{eqnarray}
F_0(t)&&=\sqrt{-2(U_0,U_0)a^{(0)}}|t|,
\ {\rm if}\  a^{(0)}>0,\ E_0=0,\\
&&=\sqrt{|a^{(0)}/E_{0}|}
\cosh[\sqrt{2(U_0,U_0)E_{0}}t],
\ {\rm if}\  a^{(0)}<0,\ E_0<0,\\
&&=\sqrt{|a^{(0)}/E_{0}|}
\sinh[\sqrt{2(U_0,U_0)E_{0}}|t|],
\ {\rm if}\  a^{(0)}>0,\ E_0<0,\\
&&=\sqrt{|a^{(0)}/E_{0}|}
\sin[\sqrt{-2(U_0,U_0)E_{0}}|t|],
\ {\rm if}\  a^{(0)}>0,\ E_0>0.
\end{eqnarray}
For $q^p$ we get
\begin{eqnarray}
&&e^{-q^p(t)}=e^{A^pt+B^p},
\ p=|S_*|+1,\ldots,|S_*|+n',\\
&&E_f=\frac{1}{2}\sum_{p=|S_*|+1}^{|S_*|+n'}
(f_p,f_p)(A^p)^2,
\end{eqnarray}
where $A^p$ and $B^p$ are arbitrary constants.

Using the following translation
\begin{eqnarray}
q^s \mapsto q^s-\ln C^s,
\end{eqnarray}
where the constants $C^s$ satisfy
\begin{eqnarray}
\sum_{s'\in S_*}C_{ss'}\ln C^{s'}=\ln[2U^2a^{(s)}], s\in S_*,
\end{eqnarray}
we get $L_{Q,T}$ and $E_{Q,T}$ in the form
\begin{eqnarray}
L_{Q,T}=\frac{1}{4U^2}L_T,\ \ \ E_{Q,T}=\frac{1}{4U^2}E_T,
\end{eqnarray}
where
\begin{eqnarray}
&&L_T=\frac{1}{2}\sum_{s,s'\in S_*}C_{ss'}\dot{q}^s\dot{q}^{s'}-V_T,
E_T=\frac{1}{2}\sum_{s,s'\in S_*}C_{ss'}\dot{q}^s\dot{q}^{s'}+V_T,\\
&&V_T=\sum_{s'\in S_*}{\rm exp}
\left[
\sum_{s'\in S_*}C_{ss'}q^{s'}
\right].
\end{eqnarray}
We notice that $L_T$ represents the Lagrangian of a Toda chain associated 
with the Lie algebra $A_{|S_*|}==sl(|S_*|+1,C)$, when the root vectors are 
put into the Chevalley basis and the coordinate describing the motion of 
the mass center is separated out. Using the result of Anderson 
\cite{Anderson}
we present the solution of the equations of motion following from $L_T$
\bea
e^{-q^s(t)} & \equiv & F_s(t)\nn\\
&=&
\sum_{r_1<\ldots<r_{m(s)}}^{|S_*|+1} v_{r_1}\cdots v_{r_{m(s)}}
\Delta^2(r_1,\ldots,r_{m(s)})e^{(w_{r_1}+\ldots +w_{r_{m(s)}})t},
\eea
where we denoted by $m(s)$ the number of element $s\in S_*$.
The $m(s)$ is assigned to $s$ in accordance with Eq.~(\ref{56}). 
$\Delta^2(r_1,\ldots,r_{m(s)})$ denotes the square of the Vandermonde 
determinant
\begin{equation}
\Delta^2(r_1,\ldots,r_m)=\prod_{r_i<r_j}\left(w_{r_i}-w_{r_j}\right)^2\; .
\end{equation}
The constants $v_r$ and $w_r$ have to satisfy the relations
\begin{equation}
\label{3.21}
\prod_{r=1}^{|S_*|+1} v_r=\Delta^{-2}(1,\ldots,|S_*|+1), 
\ \ \ \sum_{r=1}^{|S_*|+1}w_r=0\; .
\end{equation}
The energy of the Toda chain described by this solution is given by
\begin{equation}
E_T=\frac{1}{2}\sum_{r=1}^{|S_*|+1}w^2_r.
\end{equation}
Takinq into account the transformation (73), we get
\begin{eqnarray}
&&e^{-q^s(t)}\equiv \tilde{F}_s(t)=
\frac{F_s(t)}{C^s},\ \ \ s\in S_*,\\
&&E_{Q,T}=\frac{1}{4U^2}\sum_{r=1}^{|S_*|+1}w^2_r.
\end{eqnarray}
Finally, we obtain the following decomposition of the vector $x(t)$
\begin{equation}\label{84}
x(t)=
\ln[F_0(t-t_0)]\frac{U_0}{1-1/d_0}-
\sum_{s\in S_*}
\ln[\tilde{F}_s(t)]\frac{U_s}{U^2}-
\sum_{p=|S_*|+1}^{|S_*|+n'}[A^pt+B^p]f_p.
\end{equation}

We introduce the following vectors in ${\bf R}^{n+l+1}$
\begin{equation}\label{85}
a \equiv \sum_{A=0}^{n+l} a^Ae_A=
\sum_{p=|S_*|+1}^{|S_*|+n'}A^pf_p,
\ b \equiv \sum_{A=0}^{n+l} b^Ae_A=
\sum_{p=|S_*|+1}^{|S_*|+n'}B^pf_p.
\end{equation}
Their coordinates $a^A$ and $b^A$ with respect to the canonical basis 
satisfy the relations
\begin{eqnarray}
&&(a,U_0)=\sum_{A,B=0}^{n+l}G_{AB}a^AU^B_0=0,
\ (a,U_s)=\sum_{A,B=0}^{n+l}G_{AB}a^AU^B_s=0,\\
&&(b,U_0)=\sum_{A,B=0}^{n+l}G_{AB}b^AU^B_0=0,
\ (b,U_s)=\sum_{A,B=0}^{n+l}G_{AB}b^AU^B_s=0,\\
&&(a,a)=\sum_{A,B=0}^{n+l}G_{AB}a^Aa^B=2E_f,
\end{eqnarray}
$s\in S_*$.
We note that, if $n'=n+l-|S_*|=0$, we must put $a=b=0$.

Using Eq.~(\ref{84}) with Eq.~(\ref{85}), the definitions 
(46)-(48), we 
obtain the scale factors ${\rm exp}[x^i(t)]$ and the dilatonic fields 
$\varphi^{\alpha}(t)$ 
\begin{eqnarray} e^{x^0(t)}= 
[F_0(t-t_0)]^{\frac{1}{d_0-1}}
\prod_{s\in S_*}[\tilde{F}_s(t)]^
{\frac{d(I_s)}{(D-2)U^2}}
\exp\left[a^0t+b^0\right],\\
e^{x^i(t)}=
\prod_{s\in S_*}[\tilde{F}_s(t)]^
{ \frac{1}{U^2}\left( \frac{d(I_s)}{D-2}-\delta_{iI_s}\right) }
\exp\left[a^it+b^i\right],\ i=1,\ldots,n,\\
\varphi^{\alpha}(t)=
\prod_{s\in S_*}[\tilde{F}_s(t)]^
{\frac{1}{U^2}\chi_s\lambda^{\alpha}_{a_s}}
\exp\left[a^{n+1+\alpha}t+b^{n+1+\alpha}\right],\ \alpha=1,\ldots,l.
\end{eqnarray}
For the potentials $\Phi^s$ we obtain
\begin{eqnarray}
\dot{\Phi}^s(t)=Q_s
\prod_{s'\in S_*}[\tilde{F}_{s'}(t)]^ {-C_{ss'}}.
\end{eqnarray}
We remind that within this model $t$ is the time in the harmonic gauge.

\section*{Acknowledgments}
The authors are grateful to M.Rainer for useful discussions.

\end{document}